\documentclass[letterpaper,twocolumn,10pt]{article}
\usepackage{styles/usenix-2020-09}

\usepackage{algorithm}
\usepackage[noend]{algpseudocode}
\usepackage{amsmath}
\usepackage{amssymb}
\usepackage{amsfonts}
\usepackage{blindtext}
\usepackage[capitalise,nameinlink,noabbrev]{cleveref}
\usepackage{listings}
\usepackage{multirow}
\usepackage{xcolor}
\usepackage{xcolor-material}
\usepackage{tikz}
\usepackage{pgfplots}
\usepackage{pgfplotstable}
\usepackage[switch]{lineno}
\usepackage{caption}
\usepackage{subcaption}
\usepackage[inline]{enumitem}
\usepackage{booktabs}
\usepackage{pgf-pie}
\usepackage{textcomp}
\usepackage{array}
\usepackage{balance}
\usepackage{multicol}
\PassOptionsToPackage{hyphens}{url}
\usepackage{hyperref}
\usepackage{authblk}

\graphicspath{{figures/}}

\newif\ifblind
\blindfalse

\ifblind
\newcommand{\sys}[1]{\textsc{Anonymized}}
\newcommand{\az}[1]{\textsc{Anonymized}}
\else
\newcommand{\sys}[1]{{SuperBench}}
\newcommand{\az}[1]{{Azure}}
\fi

\algnewcommand\algorithmicswitch{\textbf{switch}}
\algnewcommand\algorithmiccase{\textbf{case}}
\algnewcommand\algorithmicassert{\texttt{assert}}
\algnewcommand\Assert[1]{\State \algorithmicassert(#1)}%

\algdef{SE}[SWITCH]{Switch}{EndSwitch}[1]{\algorithmicswitch\ #1\ \algorithmicdo}{\algorithmicend\ \algorithmicswitch}%
\algdef{SE}[CASE]{Case}{EndCase}[1]{\algorithmiccase\ #1}{\algorithmicend\ \algorithmiccase}%
\algtext*{EndSwitch}%
\algtext*{EndCase}%

\crefformat{footnote}{#2\footnotemark[#1]#3}
\crefname{lstlisting}{Listing}{Listings}

\pgfplotsset{compat=1.18}

\makeatletter
\long\def\ifnodedefined#1#2#3{%
    \@ifundefined{pgf@sh@ns@#1}{#3}{#2}%
}

\pgfplotsset{
    discontinuous/.style={
    scatter,
    scatter/@pre marker code/.code={
        \ifnodedefined{marker}{
            \pgfpointdiff{\pgfpointanchor{marker}{center}}%
             {\pgfpoint{0}{0}}%
             \ifdim\pgf@y>0pt
                \tikzset{options/.style={mark=*}}
                \draw [densely dashed] (marker-|0,0) -- (0,0);
                \draw plot [mark=*,mark options={fill=white}] coordinates {(marker-|0,0)};
             \else
                \tikzset{options/.style={mark=none}}
             \fi
        }{
            \tikzset{options/.style={mark=none}}        
        }
        \coordinate (marker) at (0,0);
        \begin{scope}[options]
    },
    scatter/@post marker code/.code={\end{scope}}
    }
}
\makeatother

\makeatletter
\renewcommand{\paragraph}{%
  \@startsection{paragraph}{4}%
  {\z@}{0.75ex \@plus 1ex \@minus .2ex}{-1em}%
  {\normalfont\normalsize\bfseries}%
}
\makeatother

\makeatletter
\renewcommand\AB@affilsepx{\qquad \protect\Affilfont}
\makeatother

\newcommand*\circled[1]{\tikz[baseline=(char.base)]{
            \node[shape=circle,fill,inner sep=1pt] (char) {\textcolor{white}{#1}};}}

\setlist[itemize]{itemsep=0pt,topsep=1pt,leftmargin=0.4cm}
\setlist[enumerate]{itemsep=0pt,topsep=1pt,leftmargin=0.4cm}

\begin{document}

\ifblind
\title{\sys{}: Improving Cloud AI Infrastructure Reliability with Proactive Validation}
\author{}
\else
\title{\sys{}: Improving Cloud AI Infrastructure Reliability with\\ Proactive Validation}
\author[$\dagger*$]{Yifan Xiong}
\author[$\dagger*$]{Yuting Jiang}
\author[$\dagger*$]{Ziyue Yang}
\author[$\dagger$]{Lei Qu}
\author[$\ddagger$]{Guoshuai Zhao}
\author[$\ddagger$]{Shuguang Liu}
\author[$\ddagger$]{Dong Zhong}
\author[$\ddagger$]{Boris Pinzur}
\author[$\ddagger$]{Jie Zhang}
\author[$\ddagger$]{Yang Wang}
\author[$\ddagger$]{Jithin Jose}
\author[$\ddagger$]{Hossein Pourreza}
\author[$\ddagger$]{Jeff Baxter}
\author[$\ddagger$]{Kushal Datta}
\author[$\ddagger$]{Prabhat Ram}
\author[$\ddagger$]{Luke Melton}
\author[$\ddagger$]{Joe Chau}
\author[$\dagger$]{Peng Cheng}
\author[$\dagger$]{Yongqiang Xiong}
\author[$\dagger$]{Lidong Zhou}
\affil[$\dagger$]{Microsoft Research}
\affil[$\ddagger$]{Microsoft}
\fi

\date{}
\maketitle
\ifblind\else
\def\thefootnote{*}\footnotetext{Equal contribution.}\def\thefootnote{\arabic{footnote}}
\fi

\begin{abstract}

Reliability in cloud AI infrastructure is crucial for cloud service providers, prompting the widespread use of hardware redundancies.
However, these redundancies can inadvertently lead to hidden degradation, so called ``gray failure'', for AI workloads, significantly affecting end-to-end performance and concealing performance issues, which complicates root cause analysis for failures and regressions.

We introduce \sys{}, a proactive validation system for AI infrastructure that mitigates hidden degradation caused by hardware redundancies and enhances overall reliability.
\sys{} features a comprehensive benchmark suite, capable of evaluating individual hardware components and representing most real AI workloads.
It comprises a Validator which learns benchmark criteria to clearly pinpoint defective components.
Additionally, \sys{} incorporates a Selector to balance validation time and issue-related penalties, enabling optimal timing for validation execution with a tailored subset of benchmarks.
Through testbed evaluation and simulation, we demonstrate that \sys{} can increase the mean time between incidents by up to $22.61\times$.
\sys{} has been successfully deployed in \az{} production, validating hundreds of thousands of GPUs over the last two years.

\end{abstract}

\section{Introduction}

In the past decade, the surging demand for deep learning~\cite{chatgpt, gpt4, bingchat, bard} has spurred the development of exa-scale cloud AI infrastructure, which entails significantly high costs~\cite{aisupercopmuter, sc-price, tpuv4}.
When incidents such as customer workload failures or performance regressions arise within the infrastructure, they can propagate throughout all nodes (i.e., servers) due to AI workloads' gang-scheduled and synchronization characteristics, leading to magnified penalties.
For example, during a two-month distributed training for the OPT model, Meta reported over 105 training restarts resulting from failures on more than 100 VMs in the cloud~\cite{zhang2022opt}, indicating 1.25 incidents per day and 61k GPU hours affected~\cite{logbook}.
To minimize such costly and disruptive incidents, maintaining system reliability is of paramount importance for cloud service providers.

Hardware redundancies are designed into cloud AI infrastructure for reliability purposes, including redundant GPU compute units~\cite{ampere, hopper}, GPU memory row remapping~\cite{rowremapping}, over-provisioned (under-subscribed) networking links~\cite{sc22azure}, etc.
However, component-wise redundancies can surprisingly introduce hidden degradation, so called ``gray failure''~\cite{huang2017gray} or ``fail-slow''~\cite{gunawi2018fail} in traditional cloud, and make incident patterns even more complex in AI era due to following unexpected reasons.
\begin{enumerate*}[label=(\roman*)] 
  \item
  AI workloads usually run long for weeks or months. The continuous and repetitive use of redundant components will cause them become problematic gradually with lower performance.
  For example, in \az{} A100 cluster, each InfiniBand top-of-rack (ToR) switch has multiple redundant up links.
  When part of the redundant links are broken, certain traffic patterns such as all-to-all collective communication can experience throughput regression due to congestion.
  Such redundancy introduces \textit{gradual performance degradation} rather than a binary either good or bad state for hardware components.
  \item
  This gradual degradation may not affect certain AI workloads in the early stage.
  For instance, ResNet model training can barely utilize all A100 GPU or networking resources, therefore less sensitive to trigger issues.
  Since it is difficult for cloud providers to unveil such partial redundancy loss timely with monitoring on existing workloads only, \textit{standalone tests are required} to stress the hardware to pinpoint issues.
  \item
  Redundancies can also hide performance issues to some extent.
  When broken redundancies cause AI workload regression, repairing only partial redundancies instead of all can usually resolve the current incident and recover the workload.
  For example, when multiple redundant links are broken, operators may only need to replace one link to unblock the workload in a limited time.
  However, this property \textit{decreases reliability} and increases the likelihood of future incidents when any single link goes bad.
\end{enumerate*}

As a result, although cloud providers apply hardware pre-qualification tests and timely troubleshooting for incidents, these methods are not aware of redundancies and cannot improve reliability.
First, qualification tests are performed by hardware vendors before delivery but it cannot address reliability issues that arise in cloud environment and services.
Second, troubleshooting is highly workload-dependent and can be time-consuming.
It typically only repairs the partial redundancies for recovery, rather than restoring the full redundancies of the affected and non-dominant hardware components.
As a result, the overall reliability remains low.
For example, in \az{}'s A100 clusters, the \textit{mean time between incidents} (MTBI) is $17.5$ hours, and $38.1\%$ of those incidents take more than 1-day to recover.
What's worse, troubleshooting is only performed passively when incidents are raised by customers, it cannot restore redundancies timely before affecting workloads.
However, such beforehand restoring is possible because gradual performance degradation can be detected by specific tests before incidents.

Considering the aforementioned problems, our key insight is that component-wise redundancies can surprisingly compromise the reliability of cloud AI infrastructure in unexpected ways. These redundancies can lead to incremental performance regressions, which should be identified and rectified before affecting end-to-end workloads.
Rather than relying on reactive troubleshooting, which focuses on visibly defective components, we adopt a proactive approach to validate cloud AI infrastructure components before incidents transpire.
This strategy aims to improve reliability at reasonable costs by balancing the trade-off between validation expenses and anticipated incident penalties.

To effectively minimize the MTBI, proactive validation should satisfy the following requirements:
\begin{enumerate*}[label=(\roman*)]
  \item \textit{Comprehensive}:
To detect incidents that are undetected by vendors in new clusters and only surface in customer workloads, validation must be comprehensive and encompass a wide range of AI workloads.
  \item \textit{Clear-cut}:
Given that hardware components can exhibit gradual performance degradation and measurements are prone to natural variance, it is essential to establish a clear-cut boundary between defective and normal performance.
Repetitions of the same test should also yield consistent results, rather than fluctuating between outcomes.
  \item \textit{Cost-efficient}:
Proactive validation necessitates additional measurements, which consume time. Therefore, it must be cost-efficient, ensuring that validation expenses remain significantly lower than the incident-associated penalties.
\end{enumerate*}

Nevertheless, addressing these requirements presents significant challenges.
Firstly, the sheer number of workloads and exponential node combinations result in an immense search space for all scenarios, making it impossible to encompass every aspect in the validation process.
Secondly, after performance being measured, there is no ground truth available for defective components. Identifying which components are defective is problematic, as hardware specifications cannot reliably predict workload performance. Moreover, AI hardware often exhibits substantial variations~\cite{sinha2022not}, further complicating the differentiation process.
Lastly, the validation time and MTBI can be interdependent, since fewer validated components lead to shorter times between incidents. Determining when to validate which components for optimal cost-efficiency, while achieving the longest MTBI with least measurement time, proves to be a challenging endeavor.

To address these challenges, we introduce a proactive validation system, \sys{}, featuring three key modules:
\begin{enumerate*}[label=(\roman*)]
  \item
First, we analyze workload distributions in large-scale, multi-tenant AI clusters to identify a small number of representative workloads and parameters applicable to most customers.
Based on this, we develop a comprehensive benchmark set, including end-to-end benchmarks for representative workload patterns and micro-benchmarks for individual hardware components.
  \item
A \textit{Validator} is designed to conduct a series of benchmarks on specified nodes.
To clearly distinguish between the benchmark results of defective and  functional components, the space of cumulative distribution, rather than average metrics, is used for similarity clustering to learn benchmark criteria offline.
  \item
To balance the trade-off between validation time and incident coverage, benchmarks are selected based on real-time estimations of incident probability.
A \textit{Selector} is designed to predicts node incident through a probability model and determine when to validate on which subset of benchmarks.
The selected benchmarks are executed by the Validator, generating new benchmark and defect data to periodically update criteria and probability model, allowing the system to evolve in tandem with the latest node statuses.
\end{enumerate*} 

\sys{} has been successfully deployed in one of the largest real-world AI infrastructure, \az{}, for more than 2 years, where it identifies $10.36\%$ nodes as defects, contributing to enhanced reliability.
All benchmarks in \sys{} have been open-sourced on GitHub\ifblind\footnote{\label{anon}Links are hidden for anonymization.}\else\footnote{\href{https://aka.ms/sb}{\texttt{https://github.com/microsoft/superbenchmark}}}\fi~and widely used by AI hardware vendors\ifblind\textsuperscript{\ref{anon}}\else~\cite{amdref1,amdref2}\fi~as a standard.
Simulation results demonstrate that proactive validation in \sys{} can increase MTBI by $22.61\times$ and $1.11\times$ compared to the absence of validation and full set validation without benchmark selection, while increasing user GPU hours by $4.81\times$ and reducing validation time cost by $92.07\%$, respectively.

Our contributions can be summarized as follows:
\begin{itemize}
  \item We identify the existence of gray failure in the new era of AI, which arises from AI hardware redundancies, and propose a proactive validation approach to mitigate its significant impact in cloud AI infrastructure.
  \item We observe and analyze a vast number of workload incidents, including failures and regressions, within a production cloud AI infrastructure, summarizing their sources and root causes.
  \item We establish the design goals for a proactive validation system and implement \sys{} to address these aims.
  \item We conduct extensive evaluations of \sys{} and deploy the system in a production cloud environment, demonstrating its benefits and practicality in real-world.
\end{itemize}

\section{Background and Motivation}

\subsection{Massive Incidents in AI Infrastructure}

\begin{figure}[t]
    \centering
    \resizebox{.9\linewidth}{!}{\begin{tikzpicture}[font={\Large}]


\pie[color={MaterialDeepOrange,MaterialLightBlue,MaterialCyan,MaterialGreen,MaterialLightGreen,MaterialLime,MaterialAmber,MaterialOrange}, hide number, explode=0.18]{
64.4/GPU Memory - 64.4\%,
9.2/Single Bit ECC - 9.2\%,
9.9/9.9\% - End-to-end Model Workload, 
8.3/8.3\% - InfiniBand,
2.9/2.9\% - PCIe and NVLink, 
4.1/4.1\% - GPU Temperature,
0.9/0.9\% - GPU Tensor Cores,
0.3/0.3\% - GPU CUDA Cores}

\end{tikzpicture}}
    \caption{Percentage of infrastructure incidents' sources.}
    \label{fig:issue_cat}
\end{figure}
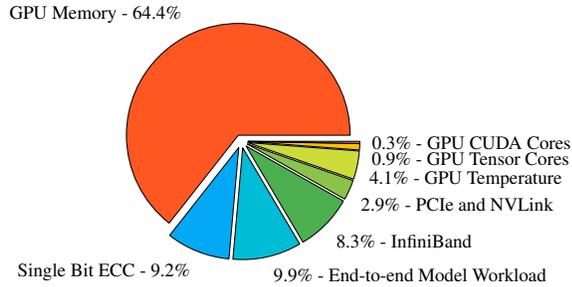

\paragraph{Incidents Summary}
In cloud AI infrastructure, we have observed that a wide range of components experience failure or performance regression. ~\cref{fig:issue_cat} illustrates that more than 8 components in GPU node can occur performance issues according to 1-month tickets in \az{}. However, even when only one component suffer issue, the real workloads will be reflected and incident ticket will be raised by customers.

\begin{figure}[t]
    \centering
    \resizebox{.9\linewidth}{!}{\usetikzlibrary{patterns}
\pgfplotsset{compat=1.18}
\begin{tikzpicture}

    \begin{axis}[
		width=\linewidth,
		height=.55\linewidth,
		xlabel={Troubleshooting Duration},
		ylabel={Incidents Percentage (\%)},
		x label style={at={(axis description cs:0.5,-0.25)},anchor=north},
		y label style={at={(axis description cs:-0.1,0.5)},anchor=south},
		ymin=0, ymax=75,
		minor y tick num = 3,
		xticklabels={,0-day,1-day,3-day,7-day,14-day,$\infty$-day},
		ticklabel style={font=\small},
		xticklabel style={rotate=30, anchor=north east},
		axis lines*=left,
		ymajorgrids=true,
		grid style=dashed,
		area style,
		]
		\addplot+[color=GoogleBlue, ybar interval, pattern=crosshatch, pattern color=GoogleBlue, line width=1pt] plot coordinates {
			(1, 13.7)
			(2, 10.3)
			(3, 3.9)
			(4, 0)
		};
		\addplot+[color=GoogleGreen, ybar interval, pattern=north east lines, pattern color=GoogleGreen, line width=1pt] plot coordinates {
			(0, 61.9)
			(1, 0)
		};
		\addplot+[color=GoogleRed, ybar interval, pattern=north west lines, pattern color=GoogleRed, line width=1pt] plot coordinates {
			(4, 10.3)
			(5, 0)
		};
		\node[above] at (axis cs: 0.5, 61.9) {61.9\%};
		\node[above] at (axis cs: 1.5, 13.7) {13.7\%};
		\node[above] at (axis cs: 2.5, 10.3) {10.3\%};
		\node[above] at (axis cs: 3.5, 3.9) {3.9\%};
		\node[above] at (axis cs: 4.5, 10.3) {10.3\%};
	\end{axis}

\end{tikzpicture}}
    \caption{Incidents troubleshooting duration distribution.}
    \label{fig:ticket_time}
\end{figure}
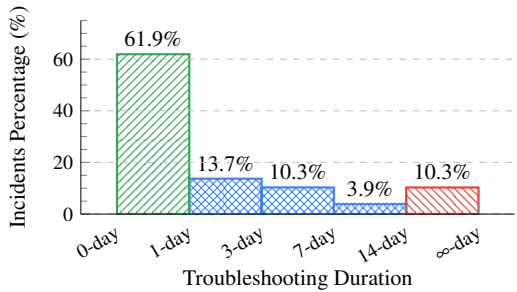

When an incident occurs in cloud AI infrastructure, it typically takes a significant amount of time for troubleshooting in order to localize the defective nodes or links for mitigation.
Meanwhile, all nodes involved in the incidents are waiting for troubleshooting, resulting in a considerable waste of GPU hours. ~\cref{fig:ticket_time} shows that 38.1\% of incidents require more than 1 day to resolve, and 10.3\% incidents take more than 2 weeks. Thus it is crucial to identify defective nodes or links timely ahead of incidents for resources saving.

Upon analyzing the related incidents, we discovered that the main reasons for high incident rate stem from three aspects: rapid-evolving AI hardware, cloud environment, and immature software.
We will explain more in the following paragraphs with real cases observed in \az{}.

\paragraph{Rapid Hardware Evolution}
AI hardware defects may not be fully identified by vendors' qualification tests, as certain issues only surface in real workloads. 
Data center GPUs, in particular, are evolving rapidly, with new releases every one or two years~\cite{ampere, hopper}.
Vendors typically only examine individual hardware components at full utilization during qualification tests, but some regressions occur exclusively under specific workload patterns.

For example, we identified a regression that only occurs in certain A100 VMs when computation and communication are executed concurrently.
Standalone computation benchmarks (e.g., GPU GEMM tests~\cite{cutlass}) or communication benchmarks (e.g., NCCL AllReduce tests~\cite{nccltests}) cannot expose this issue, as defective GPUs exhibit the same performance as non-defective ones.
The problem arises only during simultaneous computation and communication due to the overlapping traffic pattern triggering interference in L2 cache within GPU memory.

\paragraph{Cloud Environment}
Cloud infrastructure environments can introduce additional incidents, as they differ from vendors' qualification environments in terms of power, temperature, and other factors.
For example, data centers in tropical areas experience more incidents due to higher temperatures.
We observed a $35\times$ increase in defective InfiniBand links with $>10^{-12}$ bit error rate in data centers in tropical areas compared to data centers in higher latitudes, leading to significantly degraded performance for training and inference.
Another example is GPU throttling.
Even within the same data center, different racks or locations within the same rack can exhibit varying temperatures.
However, all GPUs are designed with identical cooling and heatsinks by vendors, resulting in GPUs located in warmer locations potentially experiencing thermal throttling if they cannot receive more cool air.

\paragraph{Software Immaturity}
At the application level, AI software stacks frequently iterate to co-evolve with hardware and adapt to new architectures and features.
For instance, CUDA~\cite{cudarelease} and ROCm~\cite{rocmrelease} release new versions every one or two months, while new GPUs may only support newer software versions.
Ensuring a mature and reliable software stack under such rapid evolution is difficult.
Furthermore, most AI workloads are not resilient to software or hardware failures.
Because AI training needs to synchronize tensors periodically with high frequency, any single failure or regression can quickly propagate to all involved nodes and cause workload incidents.
For example, a single node issue can cause an entire distributed training job with 100+ nodes to hang, resulting in a costly health check across all nodes by the on-call person~\cite{logbook}.

\subsection{Observations on Hardware Redundancies}

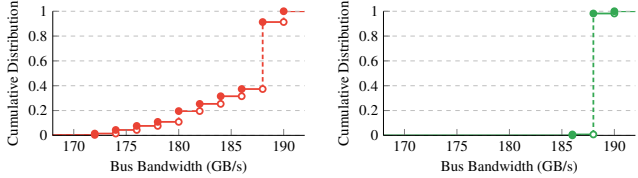
\begin{figure}[t]
    \centering
    \subfloat[Several ToR swithes have $<50\%$ of the redundant links up.]{
        \label{fig:nccl17}
        \fbox{\resizebox{.48\linewidth}{!}{\begin{tikzpicture}
    \begin{axis}[
        width=7.5cm,
        height=4.5cm,
        title={},
        xlabel={Bus Bandwidth (GB/s)}, ylabel={Cumulative Distribution},
        axis lines*=left,
        ymajorgrids=true,
        grid style=dashed,
        area style,
        jump mark left,
        xmin=168, xmax=192,
        ymin=0, ymax=1,
        every axis plot/.style={very thick},
        discontinuous,
        table/create on use/cumulative distribution/.style={
            create col/expr={\pgfmathaccuma + \thisrow{f(x)}}   
        }
    ]
    \addplot[GoogleRed] table [y=cumulative distribution]{
        x f(x)
        0 0
        168 1/276
        172 3/276
        174 8/276
        176 9/276
        178 9/276
        180 24/276
        182 16/276
        184 17/276
        186 16/276
        188 149/276
        190 24/276
        192 0
    };
    \end{axis}
\end{tikzpicture}}}
    }
    \hspace*{\fill}
    \subfloat[All ToR switches have $\ge 50\%$ of the redundant links up.]{
        \label{fig:nccl18}
        \fbox{\resizebox{.48\linewidth}{!}{\begin{tikzpicture}
    \begin{axis}[
        width=7.5cm,
        height=4.5cm,
        title={},
        xlabel={Bus Bandwidth (GB/s)}, ylabel={Cumulative Distribution},
        axis lines*=left,
        ymajorgrids=true,
        grid style=dashed,
        area style,
        jump mark left,
        xmin=168, xmax=192,
        ymin=0, ymax=1,
        every axis plot/.style={very thick},
        discontinuous,
        table/create on use/cumulative distribution/.style={
            create col/expr={\pgfmathaccuma + \thisrow{f(x)}}   
        }
    ]
    \addplot[GoogleGreen] table [y=cumulative distribution]{
        x f(x)
        0 0
        168 0
        172 0
        174 0
        176 0
        178 0
        180 0
        182 0
        184 0
        186 2/276
        188 269/276
        190 5/276
        192 0
    };
    \end{axis}
\end{tikzpicture}}}
    }
    \caption{Cumulative distribution of 2-node all-reduce bandwidth from a 24-node testbed with different redundancy ratios.}
    \label{fig:uplink}
\end{figure}

\paragraph{Over-provisioned Networking Links}
In \az{} A100 clusters, there are over-provisioned InfiniBand networks with multiple redundant links between ToR switches and aggregation switches.
However, we find that to avoid performance regression due to congestion, more than half of the redundant links must be functional.
In other words, at most half of the redundancies can be broken.
~\cref{fig:uplink} depicts a real-world case of workload regression when running on 24 nodes with 192 InfiniBand NICs in a fat-tree InfiniBand testbed.
Each ToR switch has $25\%$ redundant uplinks. When some of the edge switches have over half of the redundant links down, multiple 2-node pairs experience a significant downgrade in all-reduce bus bandwidth performance when running traffic simultaneously.
Only when operators repair the redundant links of all involved ToR switches to at least $50\%$ does the all-reduce bandwidth for all 2-node pairs return to normal performance.

\paragraph{GPU Memory Row Remapping}
A100 GPUs are equipped with redundant rows for every bank in HBM, introducing a hardware mechanism called row-remapping to replace known degraded memory cells with sparse ones in hardware and prevent the use of degraded memory~\cite{rowremapping}.
This remapping is transparent to software, with no address space changes, and the degraded memory is replaced in hardware.
~\cref{tab:row-remapping} shows that when \textgreater{}10 correctable errors are remapped in redundant rows, there is a 77.8\% higher chance of experiencing regression in end-to-end workloads compared to $1 \sim 10$ errors.

\begin{table}[t]
\centering
\caption{Row remapping impact on end-to-end workloads.}
\label{tab:row-remapping}
\begin{tabular}{|l|r|c|}
	\hline
	{\small correctable errors in row remapping} & 1 $\sim$ 10 & \textgreater{}10 \\\hline
	{\small row remapping node ratio of all nodes} & 3.19\% & 0.18\% \\\hline
	{\small regression node ratio of remapping nodes} & 5.6\% & 83.3\% \\\hline
\end{tabular}
\end{table}

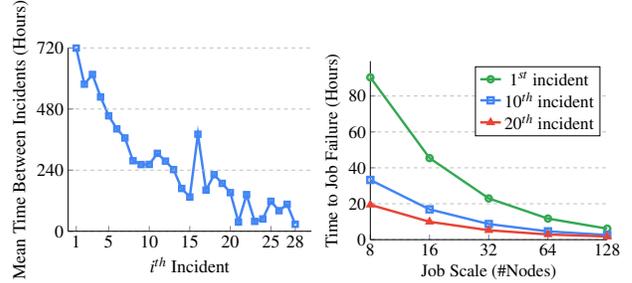
\begin{figure}[t]
    \centering
    \resizebox{.48\linewidth}{!}{\begin{tikzpicture}[font={\Large}]
    \begin{axis}[
        width=\linewidth,
        height=.8\linewidth,
        title={},
        xlabel={$i^{th}$ Incident},
        ylabel={Mean Time Between Incidents (Hours)},
        xtick={1,5,10,15,20,25,28},
        ytick={0,240,480,720},
        xmin=0, xmax=30,
        ymin=0, ymax=720,
        axis lines*=left,
        ymajorgrids=true,
        grid style=dashed,
        area style,
    ]
    \addplot[color=GoogleBlue, mark=square,line width=2pt] table {
1	719.39
2	577.66
3	615.80
4	527.41
5	453.54
6	402.03
7	366.40
8	277.02
9	262.01
10	262.57
11	306.05
12	275.56
13	242.28
14	168.17
15	134.54
16	381.09
17	161.73
18	222.21
19	188.01
20	151.71
21	36.19
22	143.83
23	38.87
24	48.42
25	117.75
26	80.77
27	105.94
28	27.73
    };
    \end{axis}
\end{tikzpicture}}
    \resizebox{.48\linewidth}{!}{\begin{tikzpicture}[font={\Large}]
	\begin{axis}[
		width=\linewidth,
        height=.8\linewidth,
        title={},
		xlabel={Job Scale (\#Nodes)}, ylabel={Time to Job Failure (Hours)},
		xmin=8, xmax=128, xmode=log, log basis x={2},
        ymin=0, 
        xticklabel={\pgfmathparse{2^\tick}\pgfmathprintnumber[fixed]{\pgfmathresult}},
		legend pos=north east,
        axis lines*=left,
		ymajorgrids=true,
		grid style=dashed,
	]
    \addplot[
		color=GoogleGreen,
		mark=o,
		mark size=2.5pt,
		line width=2pt,
	] table {
1	719.39
2	359.94
4	180.22
8	90.36
16	45.43
32	22.97
64	11.74
128	6.13
	};
	\addlegendentry{$1^{st}$ incident}
    \addplot[
		color=GoogleBlue,
		mark=square,
		mark size=2.5pt,
		line width=2pt,
	] table {
1	262.57
2	131.54
4	66.02
8	33.26
16	16.88
32	8.70
64	4.62
128	2.59
	};
	\addlegendentry{$10^{th}$ incident}
    \addplot[
		color=GoogleRed,
		mark=triangle,
		mark size=2.5pt,
		line width=2pt,
	] table {
1	151.71
2	76.10
4	38.30
8	19.41
16	9.96
32	5.24
64	2.90
128	1.75
	};
	\addlegendentry{$20^{th}$ incident}
	\end{axis}
\end{tikzpicture}}
    \caption{Left: mean time between $i$\textsuperscript{th} incidents for nodes. Right: time to failure for jobs if all nodes in the job have $i$\textsuperscript{th} incidents occurred.}
    \label{fig:time_failure}
\vspace{-1em}
\end{figure}

\paragraph{Infrastructure-wise Reliability}
Component-wise redundancy can somewhat conceal end-to-end performance issues, but it would lead to more frequent incidents over time, diminishing the reliability of the entire infrastructure.
As shown in~\cref{fig:time_failure}, we count the duration between customer-reported incidents for each node in one cluster, which has 20.7k jobs per month, then calculate the mean duration between $i$\textsuperscript{th} and $i+1$\textsuperscript{th} incidents across all nodes that have $i+1$ incidents occurred.
We find that the mean duration decreases from 719.4 hours to 151.7 hours compared between the 1\textsuperscript{st} incident and the 20\textsuperscript{th} incident.
Furthermore, we infer the time to failure for jobs at different scales, supposing all nodes in the job have the $i$\textsuperscript{th} incident occurred and share the constant failure rate.

\paragraph{Why Troubleshooting Doesn't Work}
It is hard for troubleshooting to identify the root cause of each incident and restore reliability.
In a distributed AI system, there are many components and tiers, some of which may not provide appropriate error information when an incident occurs.
For example, some incidents suffered ``uncorrectable NVLink error detected during the execution'' error reported by ML framework, but the root cause turned out to be timeout in InfiniBand network rather than broken NVLink.
A process of elimination is required to investigate each component and tier before discovering the actual cause of failure or regression.
Moreover, such workload incidents can also be closely tied to customers' applications and may not be directly related to infrastructure issues.
Additionally, due to asynchronous execution in GPUs and adaptive routing in InfiniBand networks, incidents can be non-deterministic and hard to reproduce.

Even when troubleshooting successfully locates and resolves a current failure or regression, it cannot identify all potential issues.
As AI infrastructure employs redundant resources to maintain reliability, when a failure occurs, it may indicate there are already multiple accumulated failures across various components behind the visible failure, leading to a decline in overall reliability capability.
Unfortunately, troubleshooting alone cannot fully address these types of issues.

\subsection{Challenges of Validation}
\begin{figure}[t]
    \centering
    \includegraphics[width=\linewidth]{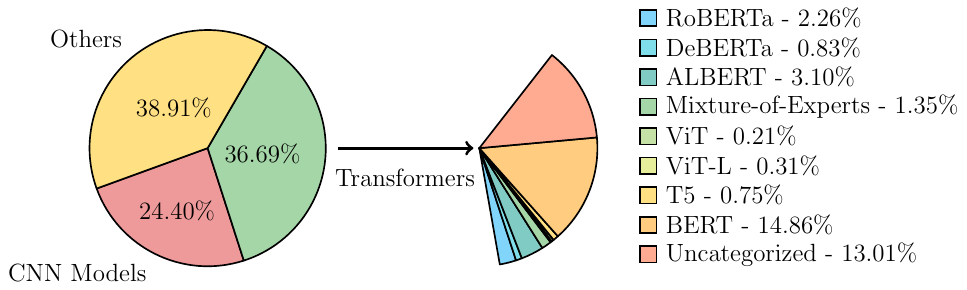}
    \caption{GPU job percentage for diverse workloads.}
    \label{fig:itp_job}
\end{figure}

Proactive validation is a practical method to tackle troubleshooting issue in AI infrastructure.
Whether validation is passed or not clearly determines whether it is the responsibility of customer workloads or hardware from the perspective of cloud providers, thereby preventing any need for customer-specific troubleshooting, such as misconfiguration.
However, there are 3 key challenges as follows:

\paragraph{Diverse Workloads}
Numerous AI models run on cloud AI infrastructure, and these AI workloads are highly diverse.
~\cref{fig:itp_job} shows the percentage of different workloads after analyzed over 56k$+$ GPU jobs in several internal clusters.
These workloads can be divided into three major categories, including Transformers, CNN, and others.
Within each category there exist tens of different models and many of them, e.g., 35.5\% of all Transformers, are hard to be identified, indicating the diversity in AI workloads.
We also analyzed the nodes with workload performance regression in this cluster in a 3-month period.
Although there were 4.7\% affected nodes in total, we found that 21.5\% of them experienced a downgrade for only one workload during troubleshooting after running 6 common workloads on the same node.
Therefore, performance regression can occur only for specific workloads, making it challenging for a validation system to ensure all customer workloads can be run with expected performance.

Distributed workloads run on multiple nodes in different scales.
The node scales and orders exponentially expands the validation scope.
For instance, all-reduce communication is commonly used during AI workloads and implemented using ring or tree algorithms.
There are $n!$ ring permutations given the same set of $n$ nodes with different orders, and different permutations utilize distinct link sets.
This causes defective links only impact certain node scale and order.
It's nearly impossible to cover all workloads with all node scales and orders in validation.

\paragraph{Lack of Ground-truth for Criteria}
Determining performance criteria is challenging, as there is no ground truth to differentiate the performance of defective components from others. Workloads typically use performance metrics like end-to-end latency or throughput. Because different workload patterns have entirely different efficiencies, the end-to-end performance cannot be directly mapped to hardware specifications (e.g., GPU FLOPS, memory bandwidth, NVLink speed). Additionally, the software stack can significantly affect these efficiencies.  Therefore, it is difficult to determine whether the measured performance is acceptable or not on a given node.

\begin{figure}[t]
    \centering
    \subfloat[LOF ($neighbors=10$) \\on VGG19 training step time]{
        \label{fig:lof}
        \includegraphics[width=.475\linewidth]{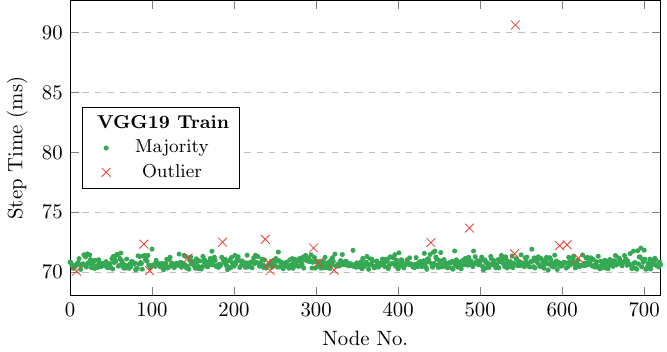}}
    \hfill
    \subfloat[One-Class SVM ($\nu=0.2$) on GPU kernel launch wall time]{
        \label{fig:OneclassSVM}
        \includegraphics[width=.475\linewidth]{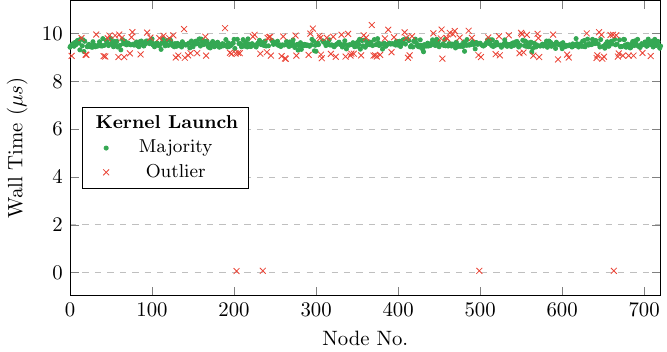}}
    \caption{Outlier detection methods for benchmark metrics.}
    \label{fig:outlier}
\end{figure}

One possible approach is adopting outlier detection to filter defects as outliers among a large number of performance data points.
However, such an approach is challenging to identify defects effectively since the distribution is diverse for different benchmarks.
The boundary between outliers and normal ones is unclear, and careful hyper-parameter tuning is
required for each benchmark on a case-by-case basis.
As shown in~\cref{fig:outlier}, the Local Outlier Factor (LOF) algorithm~\cite{breunig2000lof}, which relies on estimating data density, may mark points in expected performance but with less density as outliers.
The One-Class SVM~\cite{scholkopf2001estimating}, which relies on distances to set boundaries, may also mark false positives when data is dense within an interval.
Thus, defining the performance criteria of defective components clearly and reasonably remains challenging.

Furthermore, measurements always involve variation, including multiple runs on the same hardware or a single run on multiple hardware, which cannot be eliminated~\cite{maricq2018taming, sinha2022not}.
The variation also depends on workload types. For example, MLPerf roughly divides its benchmarks into stable vision benchmarks and higher variance benchmarks, which have $\pm 2.5\%$ and $\pm 5.0\%$ variances~\cite{mattson2020mlperf}, respectively.
Due to this, it becomes more challenging to define clear-cut criteria so that the determination of defective components remains consistent across different runs with variations.

\paragraph{Duration and Coverage Trade-off}
GPU hours in cloud AI infrastructure are expensive.
For instance, renting 1,024 A100 GPUs on Azure costs \textdollar 3.06M per month~\cite{a100price} while the same amount of TPU chips on GCP costs \textdollar 2.41M per month~\cite{tpuprice}. Fewer and shorter tests can significantly save time and cost.
However, the duration and coverage of validation can be interdependent, so less and shorter tests may reduce validation coverage and increase the risk of regression on customer workloads.
Moreover, it can be challenging to maintain high reliability over time through cost-effective validation, as hardware quality tends to decrease gradually.

\section{System Design}

\subsection{Overview}

\begin{figure*}[t]
    \centering
    \ifblind\includegraphics[width=\linewidth]{workflow}
    \else\includegraphics[width=\linewidth]{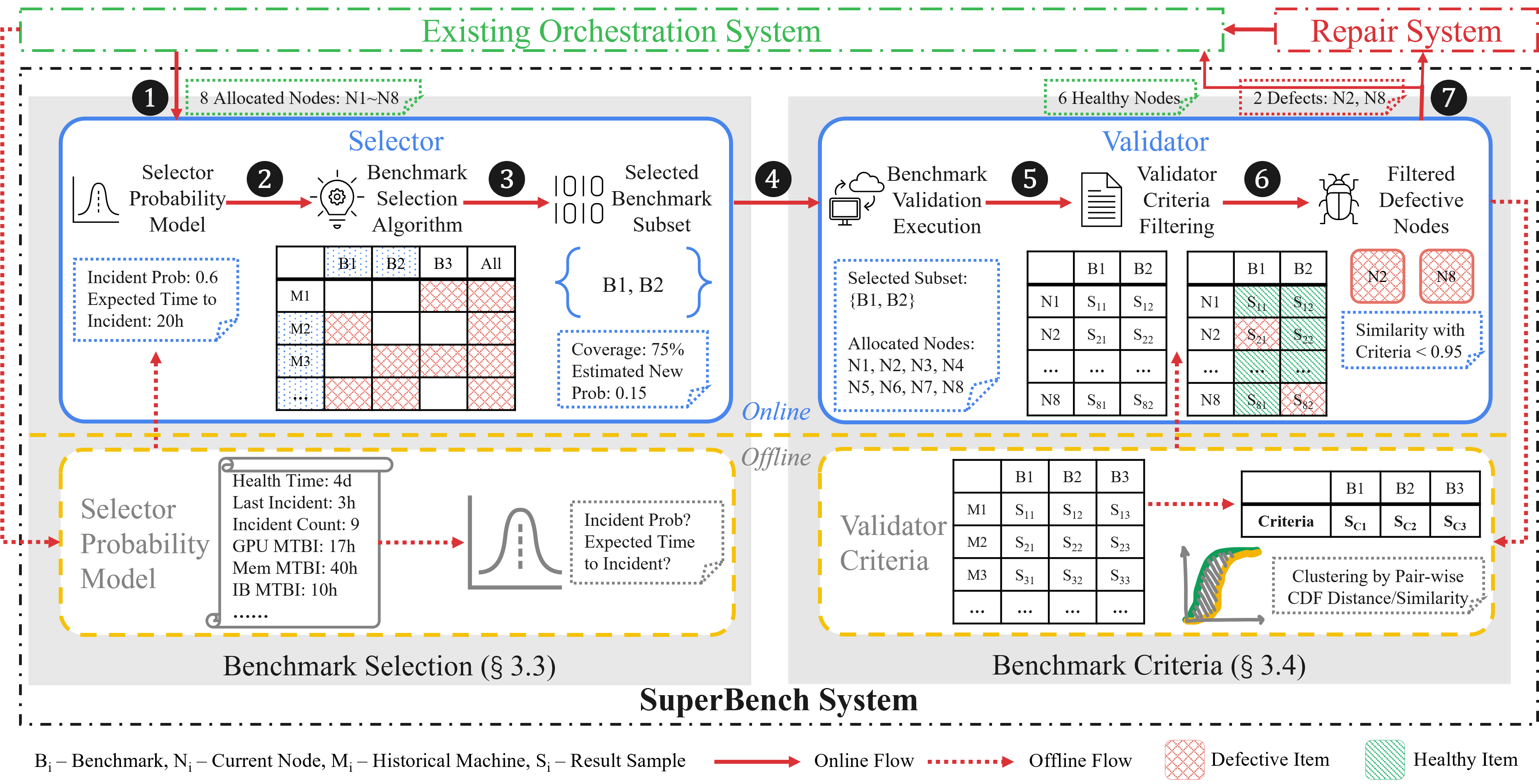}\fi
    \caption{\sys{} system architecture and an example workflow.}
    \label{fig:system}
\end{figure*}

\sys{} is proposed to tackle above challenges with the following observations:
\begin{enumerate*}[label=(\roman*)]
  \item Despite the diversity of AI workloads, a majority can be represented by a few typical workloads, while component-wise tests can cover the rest.
  \item Instead of analyzing average metrics, all measured intermediate data (or its distribution) from a single benchmark run can be employed to establish a clear-cut boundary with maximum margins between defects or normal ones.
  \item The incident probability of a node increases gradually with use. Thus, a few benchmarks may cover most potential issues if that node passed a validation involving all benchmarks in the recent past.
\end{enumerate*}

\paragraph{Principles}
The above observations guide the principles applied in the \sys{} system design:
\begin{itemize}
    \item \textit{Data-driven approaches}. Given the extensive data on defective nodes and benchmark results in cloud AI infrastructure, we utilize data-driven approaches to fit an incident probability model and formulate criteria statistically.
    \item \textit{Quick but frequent validations}. Proactive validation should not be an one-time effort. Each node should be validated quickly but frequently in line with incident probability estimations to guarantee enduring reliability.
\end{itemize}

\paragraph{Architecture}
As depicted in~\cref{fig:system}, \sys{} system comprises two primary components: a Selector and a Validator.
The Selector determines when and which benchmark to execute based on incoming events.
It offline builds a probability model to predict nodes incident probability based on their real-time statuses, and selects a benchmark subset based on a greedy algorithm for each validation event accordingly.
The Validator conducts the benchmark execution and filters defective nodes, based on execution results and offline calculated benchmark criteria.
The new node statuses and benchmark results will be continuously collected by \sys{} to periodically update the offline model and criteria.

\paragraph{Workflow}
The validation workflow is as follows.
\circled{1}
\sys{} carries out two types of validation: event-triggered and regular validation.
The former is initiated by events from orchestration system, such as the node additions, node allocations for jobs.
The latter is periodically checked by the Selector to validate existing nodes with high risks.
\circled{2}
For each validation, Selector will first obtain statuses of related nodes, like total up time, historical incident count, MTBI of different incident types, etc., and predict incident probability with an offline trained model.
\circled{3}
If the incident probability is too high for the usage, Selector will select a subset of benchmarks considering the historical benchmark coverage and validation time trade-off.
\circled{4}
The subset decision will be passed to Validator.
\circled{5}
Validator will then execute the selected subset of benchmarks on the corresponding nodes and get the results for each benchmark.
\circled{6}
The results will be compared with Validator criteria, which is offline unsupervised learned from historical benchmark results.
The node will be filtered as defective if any of its benchmark result has lower similarity with criteria than a threshold.
\circled{7}
\sys{} outputs the defective nodes for further handling.
A repair system can be employed to maintain a defective buffer with nodes to be out for repair (OFR) and a hot buffer with repaired healthy nodes. It can swap new defective nodes with healthy ones from the hot buffer and pass to orchestration system in the runtime.

\paragraph{Integration}
The proposed \sys{} validation system can be integrated into existing clusters that employ an orchestration system~\cite{vavilapalli2013apache, verma2015large, kakivaya2018service}, requiring the following interfaces supported in the control plane:
\begin{itemize}
    \item Validation Events: It requires specific events from the orchestration system to initiate validation, including
    (1) when new nodes join the cluster or cluster-wide firmware/software is upgraded,
    (2) when customer jobs are allocated to specific nodes, and
    (3) when incidents are reported by customers and particular nodes are cordoned/drained.
    \item Validation Results: \sys{} identifies the defective nodes from validation node set. The orchestration system can choose to use the rest of healthy nodes or find replaced nodes from a dedicated hot buffer.
    \item Monitored Data: The Selector probability model requires status change data from all nodes when incidents occur.
\end{itemize}

\subsection{Benchmark Set Design Choices}
\label{sec:set}

The whole validation process in \sys{} relies on a comprehensive benchmark set.
The benchmark set should be able to discover all known incidents and evolve when unseen incident occurs for the first time.
In this section, we present how it constructs a qualified set of validation benchmarks, encompassing both end-to-end benchmarks and micro-benchmarks.

\paragraph{End-to-end Benchmarks}
As the validation process aims to ensure infrastructure reliability under certain customer workloads, we first choose benchmarking the most typical end-to-end AI models to represent customer workloads and prevent potential regressions preemptively.
By examining the distribution of workloads within an internal first-party AI training platform and categorizing models based on keywords in commands and user logs, we can coorelate these workloads with several foundational models, including different CNN models~\cite{simonyan2014very, he2016deep, huang2017densely} and Transformers like BERT~\cite{vaswani2017attention} and GPT~\cite{brown2020language} models.
To optimize GPU resources and ensure training convergence, customers must also carefully configure model parameters (e.g., batch size, sequence length, etc.). We analyze parameter distributions for different foundational models and choose the most prevalent settings.

Given that workloads are continuously evolving, and current typical end-to-end benchmarks may not accurately represent future workloads, we designed the benchmark set to be extensible.
If existing benchmarks fail to detect incidents for a new workload, or private workloads run in specific clusters, the corresponding model can be added as an end-to-end benchmark and utilized for validation.
This rarely happened in our experience, except for few new dominate models.

\paragraph{Micro-benchmarks}
Rare workloads may exhibit distinct patterns when accessing specific hardware components, and the limited number of representative end-to-end benchmarks may not be able to encompass these rare workload behaviors.
Therefore, in addition to end-to-end benchmarks for typical workloads, we employ component-wise micro-benchmarks to assess individual hardware components, including CPU, memory, GPU compute units, GPU memory, NIC/HCA, disk, and their interconnections such as PCIe, xGMI, NVLink, and network links.
We leverage third-party tools accredited by hardware vendors for this purpose.

Furthermore, we derive pattern-wise micro-benchmarks in between to emulate the most common workload patterns or primitives involving several components.
For example, cuBLAS or cuDNN kernels with commonly used shapes, all-reduce or all-to-all collective communication primitives with commonly used message sizes, etc.
These patterns and primitives are profiled offline from typical workloads and enriched with regression cases during the diagnostic experience.

\subsection{Benchmark Selection}
\label{sec:selection}

Given a node set $N$, the Selector will query real-time statuses for these nodes and predict incident probability for each node through an \textit{offline probability model}.
If their joint probability $p$ of occurring an incident is higher than a threshold $p_0$, that is, the mathematical expectation of time to incident is shorter than the mathematical expectation of job duration, the Selector will select a benchmark subset for validation through an \textit{online selection algorithm}.
Otherwise the validation will be skipped to save node hours.

\paragraph{Offline Probability Model}
\label{sec:cox}
Assuming that all nodes in a large-scale cluster share the same incident probability distribution, predicting the likelihood of each node experiencing an incident during a customer workload run requires modeling the node incident distribution as a continuous function of time:
\small
\begin{equation}
P(T_{incident} \le t) = \int_{0}^{t} f(s) \,ds = F(t)
\end{equation}
\normalsize
where $f(t)$ and $F(t)$ denote the density function and cumulative distribution function of the incident time.
Traditional survival analysis~\cite{klein2003survival} suggests different statistical methods to model such probability, e.g., exponential distribution which assumes failure rate is constant over time, Weibull distribution which constrains failure rate correlated to two parameters.

Since the failure rate of GPU nodes changes over time due to the gradual degradation pattern, we need to model it without any assumptions on failure rate distribution.
The Cox-Time model~\cite{kvamme2019time}, which is an extension of the mostly used Cox proportional hazards model~\cite{cox1972regression} for time-to-event prediction in survival analysis, allows for time-varying hazard and removes the proportional assumption.
We apply the state-of-the-art Cox-Time model with neural network~\cite{kvamme2019time} for this purpose, assume node failure rate is only related to time and some covariates, and treat node statuses as covariates to failure rate, including time since the last incident, historical incident counts, mean time between incidents, etc., in different incident categories.
The Cox-Time model is trained with historical node incident data and will estimate incident probability distribution, given node's statuses and past incident history.

\paragraph{Online Selection Algorithm}
We formulate the benchmark selection problem as follows.
Given a set of benchmarks $B = \{ B_1, \dots, B_n \}$, the running time of benchmark $B_i$ as $t_i$, and a set of defective nodes $\{ M_i \}$ identified from the historical validation result data.
We assume that the full set of benchmarks shall discover all incidents, a subset of benchmarks with coverage $\mathcal{C}$ can identify the incident with probability $p \times \mathcal{C}$ hence decrease incident probability to $p \times (1-\mathcal{C})$, where the coverage is defined as percentage of defective nodes identified by the benchmark subset in history.
For example, suppose $B$ identified 10 defective nodes in the past, $B_1$ identified 2 defects $\{M_1, M_2\}$ ($\mathcal{C}=0.2$) while $B_2$ identified $\{M_2, M_3, M_4\}$ ($\mathcal{C}=0.3$).
The defect $M_2$ can be identified by either $B_1$ or $B_2$, so running the subset $\{B_1, B_2\}$ can only identify 4 defects with $\mathcal{C}=0.4$.
The goal of benchmark selection is to find a subset of benchmarks such that the new incident probability $p \times (1-\mathcal{C}) \leq p_0$ while minimizing the total benchmark time.

Since each defective node may be covered by one or more benchmarks, this problem is NP-hard and can be considered as a variation of the 0-1 knapsack problem where different items overlap.
We propose~\cref{alg:selector} to greedily select benchmarks with $\mathcal{O}(n^2)$ time complexity, where $n$ is the number of benchmarks in benchmark set $B$, instead of traversing all possible combinations with $\mathcal{O}(2^n)$ complexity, based on the sub-optimal probability decrement per time unit guideline.

\begin{algorithm}[t]
\caption{Benchmark Selection Algorithm}
\label{alg:selector}
\begin{algorithmic}[1]
\Require node set $N$, benchmark set $B$, expected prob. $p_0$
\Ensure benchmark subset $SB \subseteq B$
\Function{IncidentProb}{$N,~SB$}
\State $\mathcal{C}$ = num of defects subset $SB$ found / num of defects full set $B$ found (according to historical validation data)
\State $p = 1 - \prod_{n \in N} (1 ~-~ \operatorname*{CoxTime}_{t_0}(n))$
\State \Return $p \times (1-\mathcal{C})$
\EndFunction

\Function{BenchmarkSelection}{$N,~B$}
  \State $SB \gets \varnothing,~ p \gets$ \textsc{IncidentProb}$(N,~SB)$
  \While{$p > p_0$ \textbf{and} $SB \neq B$}
    \State $curr \gets (B_1,~0.0)$
    \For{$B_i$ \textbf{in} $B$}
      \State $\Delta p \gets p ~-~$ \textsc{IncidentProb}$(N,~SB \cup \{B_i\})$
      \If{$\Delta p / t_i > curr\left[1\right]$} $curr \gets (B_i,~\Delta p / t_i)$ \EndIf
    \EndFor
    \State $SB \cup{}= \{curr[0]\},~ p \gets$ \textsc{IncidentProb}$(N,~SB)$
  \EndWhile
  \State \Return $SB$
\EndFunction
\end{algorithmic}
\end{algorithm}

\subsection{Benchmark Criteria}
\label{sec:criteria}

Given the node set and selected benchmark subset, the Validator will execute the benchmarks on corresponding nodes to produce the benchmark results.
For the first validation during cluster build-out, the full set of benchmarks are run and \textit{offline criteria} is  unsupervised learned for each benchmark upon its results on all nodes.
For the following selective validations, the \textit{online defects filtering} will compare the benchmark result on each node to its criteria and determine the defective node if any of its result violates the criteria.
The Validator also adaptively tunes benchmark parameters for \textit{repeatability}.

\paragraph{Offline Criteria}
To clearly compare benchmark results across different runs or nodes, we first define the similarity of two measured benchmark samples.
This includes samples from micro-benchmarks, which may only have a single value as the result, or end-to-end benchmarks that track AI workload runs and record a series of performance numbers from repeated steps over a certain period.
Given two time-series benchmark samples $S_1 = \{s_{1,1}, \dots, s_{1,n}\}$ and $S_2 = \{s_{2,1}, \dots, s_{2,m}\}$, to ensure there is no asymptotic or significant regression at any time, we define the distance and similarity between two benchmark samples in their distribution space based on the empirical cumulative distribution function (CDF):
\small
\begin{equation}
\label{eq:distance}
d(S_1, S_2) = \int_{0}^{\infty} \frac{ \left| CDF_{S_1}(x) - CDF_{S_2}(x) \right|}{\operatorname*{max}( CDF_{S_1}(x), CDF_{S_2}(x) )} \,dx
\end{equation}
\begin{equation}
\label{eq:sim}
similarity(S_1, S_2) = 1 - d(S_1, S_2)
\end{equation}
\normalsize
The distance $d(S_1, S_2)$ represents the absolute integral area between their CDF curves, normalized to the $\left[0, 1\right]$ range for similarity comparison among different benchmarks.

Having a benchmark and its result samples $S_1, \dots, S_N$ from different nodes, the criteria $S_C$ is then calculated to differentiate defective result samples from normal ones.
In order to establish a clear-cut boundary, for any $S_i$ except defective samples, $S_C$ should satisfy $similarity(S_C, S_i) > \alpha$, where $\alpha$ is an empirical value set by requirements.
We therefore apply a similarity-based clustering algorithm to iteratively exclude defective samples and calculate the median sample for the rest as $S_C$, as shown in~\cref{alg:criteria}.

\begin{algorithm}[t]
\caption{Criteria Calculation Algorithm}
\label{alg:criteria}
\begin{algorithmic}[1]
\Require sample set $S = \{S_1, \dots, S_N\}$, similarity threshold $\alpha$
\Ensure criteria $S_C$ for the given benchmark
\Function{GetCentroid}{$S$}
  \State \(\triangleright\) {centroid can also be calculated by samples' mean in distribution space}
  \State $median = \operatorname*{argmax}_{i=1}^{n} \sum_{j=1}^{n} \operatorname*{similarity}(S_i, S_j)$
  \State \Return $S_{median}$
\EndFunction

\Function{CriteriaCalc}{$S, \alpha$}
  \State $Defects \gets \varnothing,~ S_C \gets$ \textsc{GetCentroid}$(S)$
  \While{$\operatorname*{min}_{S_i \in S \setminus Defects} \operatorname*{similarity}(S_C, S_i) \leq \alpha$}
    \State $Defects \gets \{ S_i \mid \operatorname*{similarity}(S_C, S_i) \leq \alpha\}$
    \State $S_C \gets$ \textsc{GetCentroid}$(S \setminus Defects)$
  \EndWhile
  \State \Return $S_C$
\EndFunction
\end{algorithmic}
\end{algorithm}

\paragraph{Online Defects Filtering}
The Validator leverages benchmark similarity between the measured benchmark result and its criteria to determine whether the node should be considered defective or not.
Since validation only requires testing whether the result on given node exhibits worse performance (i.e., lower throughput or larger latency) than the normal performance to identify defects, we compare the one-direction distance derived from~\cref{eq:distance} between the runtime observed result sample $S_{Obs}$ and offline criteria $S_C$ for throughput-like metrics (elsewise replace max with min):
\small
\begin{equation}
d_{1-side}(S_{Obs}, S_C) = \int_{0}^{\infty} \frac{ \operatorname*{max}( 0, CDF_{S_{Obs}}(x) - CDF_{S_C}(x) )}{\operatorname*{max}( CDF_{S_{Obs}}(x),~ CDF_{S_C}(x) )} \,dx
\end{equation}
\normalsize
Similarity is then calculated by $1-d_{1-side}$ and the same empirical threshold $\alpha$ is used to filter out under-performant nodes, which will be processed by the repair system next.

\paragraph{Repeatability}
Due to the natural variations when running the same workload across different runs or nodes, particularly in AI software/hardware, it is essential to ensure that validation benchmarks are minimally affected by such variations so that the measured samples can accurately represent the underlying distribution of both hardware performance and variation.
We define the \textit{repeatability} metric as \textit{``the arithmetic mean of pairwise similarities from $N$ different nodes or runs''} to ascertain whether benchmark settings are acceptable within the empirical similarity threshold $\alpha$.

To effectively validate nodes, the \textit{repeatability} metric must be maximized.
However, out-of-the-box benchmarks from public tools or hardware vendors may exhibit large variations, making them unsuitable for validation requirements.
We utilize existing benchmarks as a starting point and adhere to the guidelines below to enhance their repeatability:
\begin{enumerate}
  \item Decouple different hardware components to distinguish various sources of variation. For example, binding processes to local cores and memory to avoid random remote memory access when validating GPU, loading training data from memory to eliminate disk access variation, etc.
  \item Adaptively search for benchmark parameters to reduce benchmark duration for the given hardware/software combination, validating nodes in the shortest possible time while maintaining repeatability within the threshold.
  \item After firmware/driver updates, re-tune and re-evaluate the repeatability in case it deteriorates on newer versions.
\end{enumerate}

\section{Implementation}

\paragraph{Benchmark Set}
~\cref{tab:benchmark} shows the full benchmark set chosen for \sys{}.
Each benchmark has pre-defined and configurable parameters, such as GEMM shapes, message sizes, model batch sizes, etc., as well as randomly generated input data.
During each validation, the Selector selects a subset of benchmarks, and the Validator subsequently executes those benchmarks in two phases in sequence: the single-node phase and the multiple-node phase.
In each phase, the selected benchmarks are performed in a bottom-up manner.
Micro-benchmarks are conducted first to validate individual components, followed by end-to-end benchmarks to simulate customer workloads.
Defective nodes are removed after each phase to ensure they do not affect subsequent benchmarks.

\begin{table}[t]
    \centering
    \caption{Full benchmark set in \sys{}.}
    \label{tab:benchmark}
    \footnotesize
    \setlength{\leftmargini}{0.4cm}
    \setlength{\leftmarginii}{0.4cm}
    \begin{tabular}{| m{0.9cm} | m{3.55cm} | m{2.7cm} |}
        \hline
        & Micro benchmarks & End-to-end \newline benchmarks \\
        \hline
        Single \newline Node \newline Phase &
        Computation
        \begin{itemize}[noitemsep]
            \item GPU kernel launch
            \item GPU GEMM~\cite{cutlass, rocblas}
            \item cuBLAS kernels
            \item cuDNN kernels
            \item GPU burn
        \end{itemize}
        Communication
        \begin{itemize}[noitemsep]
            \item CPU latency~\cite{mlc}
            \item GPU H2D/D2H bandwidth
            \item GPU copy bandwidth
            \item NVLink all-reduce
            \item IB HCA loopback~\cite{perftest}
            \item IB single-node all-reduce
        \end{itemize}
        Comput./Comm. Overlap
        \begin{itemize}[noitemsep]
            \item MatMul/all-reduce overlap
            \item Sharding MatMul
        \end{itemize}
        Disk IO
        \begin{itemize}[noitemsep]
            \item FIO rand/seq read/write~\cite{fio}
        \end{itemize} &
        Multi-GPU training
        \begin{itemize}[noitemsep]
            \item CNN models
                \begin{itemize}[nosep,leftmargin=0.35cm]
                    \item ResNet~\cite{he2016deep} 50/101/152
                    \item DenseNet~\cite{huang2017densely} 169/201
                    \item VGG~\cite{simonyan2014very} 11/13/16/19
                \end{itemize}
            \item RNN models
                \begin{itemize}[nosep,leftmargin=0.35cm]
                    \item LSTM
                \end{itemize}
            \item Transformers
                \begin{itemize}[nosep,leftmargin=0.35cm]
                    \item BERT~\cite{vaswani2017attention} base/large
                    \item GPT-2~\cite{brown2020language} {} small/large
                \end{itemize}
            \item Long-running stress
                \begin{itemize}[nosep,leftmargin=0.35cm]
                    \item GPT-2 large
                \end{itemize}
        \end{itemize} \\
        \hline
        Multiple \newline Node \newline Phase &
        Networking
        \begin{itemize}[noitemsep]
            \item All pair RDMA verbs
            \item GPU collective \newline communication~\cite{nccltests, rccltests}
                \begin{itemize}[nosep,leftmargin=0.35cm]
                    \item all-reduce
                    \item all-gather
                    \item all-to-all
                \end{itemize}
        \end{itemize} &
        Multi-node training
        \begin{itemize}[nosep]
            \item CNN models
            \item RNN models
            \item Transformers
            \item Long-running stress
        \end{itemize}
        (same as above) \\
        \hline
    \end{tabular}
\end{table}

\paragraph{Networking Validation}
For the distributed microbenchmark which performs pairwise scans for RDMA verbs to ensure the networking quality of InfiniBand or RoCEv2 during large-scale AI training, we propose the following two algorithms to reduce validation time:

\textbf{Full Scan in $\mathcal{O}(n)$}
The pairwise scans for all nodes is to ensure that the networking bandwidth between any two nodes are expected so that AI workloads with all-to-all traffic (e.g., Mixture-of-Experts models) won't have regression.
The key idea is to simultaneously run network tests for all nodes, which can avoid traffic collision by leveraging the Clos network~\cite{clos1953study} property.
The problem can be formulated as follows.
Given a network with ${N}$ NICs, suppose ${N} \in 2~\mathbb{N}$ and pairs are symmetric, schedule all possible $\frac{{N} \cdot ({N} - 1)}{2}$ NIC pairs into ${N} - 1$ rounds.
In each round, exactly $\frac{{N}}{2}$ pairs will run pairwise benchmarks (e.g., GPU-direct RDMA write~\cite{perftest}, NCCL/RCCL~\cite{nccltests, rccltests} 2-node all-reduce) without NIC intersection.
We address this problem by leveraging the circle method in round-robin tournaments~\cite{kirkman1847problem} as the scheduling algorithm.

\textbf{Quick Scan in $\mathcal{O}(1)$}
To eliminate the increasing time required for large-scale networking validation, we further propose topology-aware networking validation, which scans network links within a fixed duration, irrespective of the network scale or node number.
We select node pairs such that the distance (number of hops) between each pair is 2-hop (nodes under the same ToR switch), 4-hop (nodes under the same aggregation switch), 6-hop (nodes across the core switch), and so on.
For each hop number, we include all nodes without node overlap, ensuring that every node is included once and only once for the given hop.
For a $k$-tier fat-tree network topology, only $k$ rounds are needed for any scale.
In each round, we run the selected networking benchmarks for each hop number.

\paragraph{Benchmark Parameter Searching}
For end-to-end benchmarks, validation only needs to measure a number of iterations after performance becomes stable, rather than running into model convergence or completion.
We offline search for such optimal warmup and measurement steps to reduce the validation time.
Since model structures and hyper-parameters like batch size and sequence length have already been determined based on representative workloads, the search focuses on the number of warmup steps $w$ and measurement steps $n$.
It aims to minimize the total step number, i.e., benchmark running time, while maximizing the repeatability across nodes.
This process is performed offline.

We formulate the problem as follows: Assuming the benchmark runs $K$ steps periodically and has a sequence of throughput numbers ${t_1, \dots, t_K}$, the objective is to search for parameters $w, n \in \left[1, K\right]$ such that $n$ is minimized and sub-sequence ${t_w, \dots, t_n}$ is self-similar within the similarity threshold $\alpha$ (same as the threshold used in~\cref{sec:criteria}).
First, we calculate the period of the cycle $p$ across all $K$ steps using classical seasonal decomposition by moving averages~\cite{seabold2010statsmodels} and divide them into $\frac{K}{p}$ cycles.
Next, we traverse from the beginning to calculate the pairwise similarity of all cycles, stop when a certain number of continuous similar periods are found.
$w$ and $n$ are set to the beginning and end of those continuous periods.
Finally, $(w_i, n_i)$ from different nodes are traversed to select the one that maximizes the average of pairwise similarity across all nodes.

\section{Evaluation}

\subsection{Experiment Settings}

\sys{} has been deployed in \az{} production environments since Day-0 when new AI SKUs were introduced, so it is hard to perform an apple-to-apple comparison between the complete \sys{} system and the original system without validation.
Consequently, we divide our experiment settings into two parts: First, we gather node incident statistics and traces from on-premise clusters and use this data to simulate node incidents with and without validation, assessing the efficacy of the offline-trained incident probability model and the selection algorithm in the Selector.
Second, we execute the full set of benchmarks on production clusters and two testbeds to collect benchmark results for criteria evaluation in the Validator.

\paragraph{Node Incident Trace}
We collect a 4-month period node incident events, including all node failures and customer tickets, from different internal on-premise GPU clusters with 1k nodes in total.
These clusters share the same hardware as cloud VMs but never run validation.
Each incident event has timestamps of when the event started and ended, incident reasons, and involved hardware components.
We use this trace to fit the probability model with each node's status during simulation.

\paragraph{Node Allocation Request Trace}
The internal on-premise GPU cluster deploys a container orchestration system where users can submit GPU training jobs to run.
We also collect allocation requests within the same clusters, including the requested node number, submission timestamp, and duration, to simulate node allocation requests for benchmark selection.

\paragraph{Cluster Benchmark Dataset}
We construct a benchmark dataset using the results obtained from the full set of benchmarks during the production cluster build-out. This dataset comprises 3k$+$ A100 VMs and 24 benchmarks with 2,441 metrics on each VM.
We use this dataset to build criteria, label defective nodes, and calculate the defect coverage of each benchmark for the benchmark selection algorithm.

\paragraph{Testbed}
We utilize the following internal GPU testbeds:
\begin{itemize}
    \item $144\times$ MI250X VMs. Each VM has $8\times$ AMD Instinct MI250X 120 GB GPUs~\cite{mi200} and $96\times$ 2nd-Gen AMD Epyc cores. GPUs are inter-connected by xGMI links, while VMs are connected by $8\times$ 200 Gbps HDR InfiniBand.
    \item $64\times$ H100 VMs. Each VM has $8\times$ NVIDIA H100 80 GB SXM GPUs~\cite{hopper} and $96\times$ 4th-Gen Intel Xeon cores. GPUs are inter-connected by NVLink and NVSwitch, while VMs are connected by $8\times$ 400 Gbps NDR InfiniBand.
\end{itemize}

\paragraph{Setup}
We deploy \sys{} to the testbeds and run all benchmarks with the Validator.
The environment for validation uses the exact same settings as production clusters, including firmware versions, hypervisor settings, BIOS configurations, etc.
We fixed all software versions by using a dedicated VHD image for VM and a Docker image for validation execution.
The VHD image is based on Ubuntu 22.04\ifblind\else~\cite{azhpcimage}\fi, while the Docker image is based on NVIDIA NGC container image 20.12~\cite{nvcr}.
Both images include the necessary GPU/OFED drivers and related libraries.
For the benchmark selection, we leverage PyCox~\cite{kvamme2019time} to build and train the probability model.

\subsection{Benchmark Selection Evaluation}
\label{subsec:selector-eval}

\paragraph{Offline Probability Model}
We evaluate whether the Cox-Time model in~\cref{sec:selection} can effectively use its fitted probability distribution to predict the \textit{time before next incident} (TBNI) given node statuses, including total up time, historic incident counts and MTBI in different categories, etc.
These status numbers are concatenated into a feature vector as the model's input, while the expected TBNI is used as prediction output.
Failure to predict TBNI for the given nodes will result in inappropriate benchmark selections hence higher MTBI.
We extract 46,808 node status samples and corresponding TBNI from the node incident trace.
80\% of the samples are used for model training and 20\% are used for evaluation.

We establish three baselines for comparison:
\begin{itemize}
    \item Exponential Distribution $S(t)=e^{-\lambda t}$~\cite{birolini2007reliability}. It assumes the incident rate $\lambda$ is constant across different node statuses.
    \item Exponential Distribution per Incident Count. It assumes the incident rate is simply related to historical incident count, as informed by~\cref{fig:time_failure}, and constructs exponential distributions for node statues with the same incident count.
    \item Exponential Distribution per Hour. It assumes the incident rate is related to current up time. The incident rate for each hour $H$ is calculated by dividing the number of samples with at least an $H$-hour life by the total number of samples.
\end{itemize}

We compare the accuracy of different probability models.
For each sample, which contains one single incident, the prediction accuracy is calculated by $\frac{\left\| \left[prediction\right] - \left[TBNI\right] \right\|}{2400}$, where the prediction is capped at the trace length of 2,400 hours (100 days) to make accuracy $\leq 100\%$.
The model accuracy is defined as the average of its accuracy on all test samples.

The results are shown in~\cref{tab:cox-time-accuracy}.
The first and third models yield the same accuracy since they both predict $>2,400$ hours TBNI for all samples.
The Cox-Time model, which considers multiple node statuse as covariates, achieves a high accuracy of 93.1\%, significantly outperforming the other baseline models.
This result also highlights the important relationship between node statuses and incident probability distribution.

\begin{table}[t]
\centering
\small
\caption{Accuracy of different probability models.}
\label{tab:cox-time-accuracy}
\begin{tabular}{cc}
    \toprule
    \textbf{Model} & \textbf{Accuracy} \\\midrule
    Exponential Distribution & 75.12\% \\
    Exponential Distribution per Incident Count & 63.03\% \\
    Exponential Distribution per Hour & 75.12\% \\
    Cox-Time Model & \textbf{93.13\%} \\\bottomrule
\end{tabular}
\end{table}

\paragraph{Online Selection Simulation}
To understand whether the benchmark selection can proactively identify defective nodes and save node hours, we simulate customer jobs in a cluster and evaluate whether the benchmark selection reduces incidents and improves utilization.
The simulation will mark nodes with incidents under certain categories.
Note that we assume the full set of benchmarks can always discover all incidents and whether the selected subset can proactively prevent the incident is decided by coverage from historical validation data.
We simulate 720 hours (30 days) of cluster activities. The detailed simulation process is described as follows:
\begin{enumerate}
\item Simulator sets up FIFO queues for both job and node, and employ \textit{stressed replay} based on the allocation request trace to schedule jobs to the nodes in a best-effort manner.
\item For current allocated nodes, simulator will sample the next incident time and an incident category for each node based on the probability distribution trained on incident trace.
\item Selector selects a subset of benchmarks by~\cref{alg:selector}.
\item Simulator calculates the probability on whether the selected subset can identify simulated incidents based on benchmark coverage instead of running actual benchmarks.
\item If the Selector filters defects before job running, related defective nodes will be sent to repair, and the rest nodes and the job will be pushed to the end of respective queues.
\item If the Selector fails to predict defects that cause incidents during job running, a new benchmark selection will be performed based on updated node statuses. The job will be pushed to the queue rear and continue where it lefts off.
\item For the node repairing, the duration is 1.5 days (expectancy of ticket time in~\cref{fig:ticket_time}) for no validation case to troubleshoot potential problematic nodes, and 1 hour for validation case to replace defects with healthy ones in hot buffer according to the empirical experience.
\end{enumerate}

Apart from the~\cref{alg:selector}, we establish three baselines for comparison:
(1) absence of validation, no validation and nodes are sent for repair upon each incident,
(2) full set validation, where nodes run validation with the full set of benchmarks upon each job allocation and each incident, and
(3) ideal baseline, where all nodes are healthy forever and no incidents occur.

\begin{figure}[t]
    \centering
    \includegraphics[width=\linewidth]{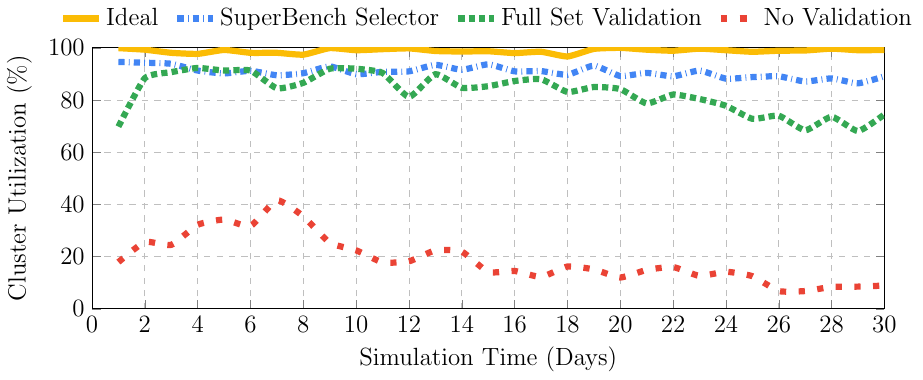}
    \caption{Simulated average node utilization with different benchmark selection policies within 30 days.}
    \label{fig:simulation_utils}
\end{figure}

\begin{table}[t]
\centering
\footnotesize
\caption{Simulated average node validation time and MTBI with different benchmark selection policies in 30-day.}
\label{tab:simulation-mtbi}
\begin{tabular}{c|ccc}
	\hline
	\textbf{Policy} & Absence & Full Set & \sys{} Selector \\\hline
	\textbf{Validation Time (h)} & 0 & 100.40 & \textbf{7.96} \\\hline
	\textbf{MTBI (h)} & 11.59 & 236.26 & \textbf{262.05} \\\hline
\end{tabular}
\end{table}

We evaluate the effectiveness of different policies, focusing on three metrics: average node utilization, average node validation time cost, and node MTBI of a cluster.
For each node, its utilization is calculated by dividing its up time by total time, its validation time is the sum of the durations of all validations, while its MTBI is calculated by dividing its up time by the number of incidents that occurred within it.

The average node utilization results are presented in~\cref{fig:simulation_utils}. \sys{} Selector with benchmark selection algorithm achieves a high cluster utilization of $90.70\%$, improving the no validation baseline by $4.81\times$ and the full set baseline by $1.09\times$.
As depicted in~\cref{tab:simulation-mtbi}, the Selector reduces $92.07\%$ validation time cost per node compared to the full set baseline, which is irrelevant to simulation duration since validation is performed regularly in production.
In terms of MTBI, the Selector achieves a high $262.05$ hours, enhancing the no validation baseline by $22.61\times$ and the full set baseline by $1.11\times$.
The Selector experiences an average of 4.79 incidents per node, 0.16 higher than full set baseline.
Despite more incidents, it benefits from reduced down time due to selective validation, resulting in longer up time and better MTBI.

\subsection{Benchmark Criteria Evaluation}
\label{evaluation-validator}

\paragraph{Criteria and Defects Filtering}

\begin{figure}[t]
    \centering
    \ifblind\includegraphics[width=\linewidth]{criteria}
    \else\includegraphics[width=\linewidth]{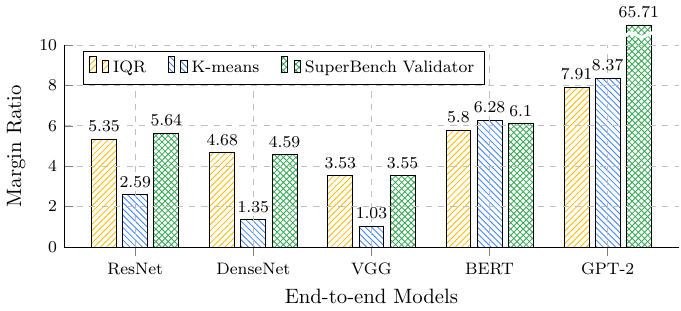}\fi
    \caption{Margin ratios of different criteria methods.}
    \label{fig:identify-defects}
\end{figure}

To evaluate benchmark criteria, we collect a series of step throughput for all end-to-end benchmarks on 144 MI250X VMs and generate the criteria using different methods to compare their effectiveness.
Note that there is no ground-truth on which nodes are defective.
Any nodes with failures or performance regressions are defined as defects by definition.
To compare whether the criteria can maximize the margin between healthy and defective nodes, we define the \textit{Margin Ratio} metric as $\frac{\operatorname*{min}_{~i \in defective} d(S_i,~S_C)}{\operatorname*{max}_{~j \in healthy~} d(S_j,~S_C)}$.

Apart from the proposed method, we establish two baselines with typical outlier detection methods, including the interquartile range (IQR)~\cite{dekking2005modern} and $k$-means~\cite{hartigan1979algorithm}.
For the proposed method, we use criteria $S_C$ calculated by~\cref{alg:criteria} and choose $\alpha=0.95$ as the similarity threshold.
For IQR, we use average throughput of each sample to calculate the lower quartile $Q_1$ and upper quartile $Q_3$~\cite{dekking2005modern}.
The $S_C$ for IQR is then calculated as the median of samples higher than $Q_1 - 1.5 \cdot (Q_3 - Q_1)$.
For $k$-means, we use the default Euclidean distance as the benchmark samples distance in the $k$-means clustering algorithm and set the cluster number $k=2$.
$S_C$ is set to the average of all samples in the majority cluster.

As shown in~\cref{fig:identify-defects}, out of a total of 5 models, IQR and $k$-means both have 4 models worse than the proposed~\cref{alg:criteria}, which can achieve relatively better margin ratios across different benchmarks compared with baseline methods.
For GPT-2, both baselines classify nodes with marginal performance as defective, which subsequently decreases the numerator in the margin ratio calculation, leading to an unclear boundary.
On the contrary, the proposed method identifies them as healthy to maximize the margin between healthy and defective nodes, resulting in significantly larger margin ratio.

\paragraph{Repeatability}
We evaluate the effectiveness of the proposed adaptive benchmark parameter searching by comparing its repeatability to a baseline with sufficiently long and fixed step parameters, i.e., 72 warmup and 3,072 measurement steps.
We execute end-to-end benchmarks on 64 H100 VMs and collect the warmup and measurement step numbers for each run.
The repeatability metric is measured as the arithmetic mean of similarity scores between each sample and criteria.
As shown in~\cref{tab:parameters-tuning}, repeatability regression of the proposed method (tuned parameters) is less than 1\% compared to the baseline (fixed parameters), while saving $67.5\% \sim 78.3\%$ validation time cost for all models.

\begin{table}[t]
\centering
\scriptsize
\caption{Repeatability after benchmark parameters tuned.}
\label{tab:parameters-tuning}
\begin{tabular}{c|ccc}
    \toprule
    \textbf{End-to-end} & \multicolumn{2}{c}{\textbf{Repeatability (FP32 / FP16)}} & \textbf{Time Saving} \\
    \textbf{Models} & \textbf{\scriptsize Fixed Parameters} & \textbf{\scriptsize Tuned Parameters} & \textbf{FP32 / FP16} \\
    \midrule
    ResNet & 98.70\% / 97.52\% & 98.70\% / 97.55\% & 73.96\% / 78.30\% \\
    DenseNet & 99.13\% / 98.82\% & 99.12\% / 98.76\% & 73.96\% / 73.96\% \\
    VGG & 98.63\% / 97.38\% & 98.62\% / 97.51\% & 75.59\% / 70.70\% \\
    LSTM & 99.51\% / 98.11\% & 99.52\% / 98.11\% & 73.96\% / 73.96\% \\
    BERT & 99.62\% / 99.44\% & 99.62\% / 99.44\% & 73.96\% / 77.21\% \\
    GPT-2 & 99.51\% / 99.37\% & 99.48\% / 99.36\% & 73.96\% / 67.45\% \\
    \bottomrule
\end{tabular}
\end{table}

\subsection{Deployment in Cloud}

\begin{table}[t]
\centering
\footnotesize
\caption{Effectiveness and repeatability in real deployment.}
\label{tab:kusto}
\begin{tabular}{c|cc}
    \toprule
    \textbf{Benchmark} & \textbf{Repeatability} & \textbf{\# Defects / \# Total} \\
    \midrule
    IB HCA loopback & 99.96\% & 6.04\% \\
    H2D/D2H bandwidth & 99.68\% & 2.03\% \\
    BERT models & 99.39\% & 1.59\% \\
    CPU latency & 99.60\% & 1.33\% \\
    IB single-node all-reduce & 99.90\% & 1.10\% \\
    ResNet models & 99.21\% & 0.73\% \\
    GPT-2 models & 99.19\% & 0.53\% \\
    LSTM models & 98.66\% & 0.46\% \\
    DenseNet models & 97.70\% & 0.40\% \\
    MatMul/all-reduce overlap & 97.88\% & 0.33\% \\
    NVLink all-reduce & 99.89\% & 0.30\% \\
    GPU GEMM & 99.44\% & 0.23\% \\
    \bottomrule
\end{tabular}
\end{table}

We have deployed \sys{} in \az{} production environment for two years, where it acts as a quality gate to identify most defective nodes before delivering GPU clusters to public cloud services.
We collect the benchmark dataset during the cluster build-out phase from 24k$+$ A100 GPUs (3k$+$ VMs) in 90 days and demonstrate its practicality.

\paragraph{Benchmark Effectiveness}
We generate criteria for the dataset and use it to filter out defective nodes with failure or regression, and calculate the percentage of defects identified by each benchmark, including micro-benchmarks and end-to-end benchmarks, as shown in~\cref{tab:kusto}.
The defects percentages are sorted in reverse order, and each defective node may be filtered by one or several benchmarks.
In total, $10.36\%$ nodes are filtered as defects without duplication.

\paragraph{Repeatability at Scale}
We also analyze the repeatability of those effective micro-benchmarks and end-to-end benchmarks among healthy nodes after filtering out defective ones.
~\cref{tab:kusto} also shows the repeatability on around 3k nodes, where most benchmarks have it greater than $99.0\%$, while all effective benchmarks have greater than $97.5\%$ repeatability.

\section{Related Works}

\paragraph{Gray Failure in Cloud}
There exist related works that explore the concept of gradual performance degradation under different terms, as opposite to fail-stop failures.
For example, ``fail-stutter''~\cite{arpaci2001fail} model is proposed to address the issue of the same components behaving differently.
Do et al.~\cite{do2013limplock} and Gunawi et al.~\cite{gunawi2014bugs} evaluate and study the ``limping'' hardware in cloud which exhibits significant degradation.
Huang et al.~\cite{huang2017gray} discuss the ``gray failure'' problem in cloud systems and its differential observability traits.
Lou et al.~\cite{lou2020understanding} evaluate the ``partial failures'' in software.
Gunawi et al.~\cite{gunawi2018fail} analyze a variety of ``fail-slow'' hardware incidents within production systems, providing valuable observations and suggestions.

In the new era of AI, computing resources become even more powerful~\cite{top500} and such gradual degradation patterns cause more significant consequences due to gang-scheduled and synchronized AI workloads.
We study this degradation pattern in the GPU-based cloud at scale and correlate it with the redundancies in hardware.
Instead of diagnosing issues as they occur, we advocate for a proactive validation approach that regularly validates all components to restore redundancies, consequently mitigating the impacts of such degradation. 

\paragraph{Elastic Training and Fault-tolerant AI}
In synchronized AI workloads, one slow GPU can make all other GPUs wait and stall the entire training process.
Popular deep learning frameworks support elastic distributed training to address this issue, such as Torch Elastic~\cite{torchelastic} and Horovod Elastic~\cite{horovodelastic}, which will scale down nodes when there are hardware failures and scale up after recovery.
In addition to ML-agnostic solutions, there are also algorithmic aware fault-tolerant techniques~\cite{qiao2018litz, qiao2019fault} for AI training to bypass faulty nodes.

However, these elastic and fault-tolerant techniques may also introduce non-determinism for workloads and affect training convergence or accuracy~\cite{li2022easyscale}, which is not transparent to customers.
In contrast, the validation system we proposed aims to reduce the faulty hardware in infrastructure from the cloud service provider's perspective with no assumptions on customer workloads.

\paragraph{Performance Benchmarking}
Performance benchmarking is used to measure which computer system can complete tasks more quickly across different types.
For traditional CPU-based systems, SPEC~\cite{dixit1993overview} and LINPACK~\cite{dongarra1987linpack} are widely-used to rank supercomputers.
For AI systems, particularly GPU-based, benchmarking also shares the same goal of measuring which AI system is faster.
DeepBench~\cite{deepbench} compares typical operations in deep learning, while DAWNBench~\cite{coleman2017dawnbench} and MLPerf~\cite{mattson2020mlperf} consider end-to-end training or inference performance and measure time to convergence as metrics.

All the above benchmarks for benchmarking target ranking or competition for peak performance of different types of systems.
In contrast, the performance validation we proposed in this work aims to identify defects across reproductions of the same system, which has entirely different requirements.

\section{Conclusion}

Inherent redundancies in cloud AI infrastructure can give rise to a unique degradation pattern for AI workloads, leading to diminished reliability.
In this paper, we delved into incidents associated with this concealed degradation pattern and addressed it through the design of a proactive validation system, \sys{}.
The system comprises a comprehensive benchmark suite, a Selector designed to efficiently trade-off validation time and coverage, and a Validator which executes benchmarks and filters defects with clear-cut criteria.
The system also iterates and evolves with the rich data gathered by itself.
Our evaluation in testbed and simulation demonstrates its ability to increase the MTBI by $22.61\times$.
Moreover, \sys{} has showcased its practicality in a real-world cloud deployment.

\section*{Acknowledgments}

We sincerely thank our colleagues from Systems and Networking Research Group at Microsoft Research Asia and Azure HPC/AI team for their early and valuable feedback on this work.
We would also like to thank the anonymous reviewers for their detailed and valuable feedback on this paper.

\bibliographystyle{plain}
\bibliography{reference}

\end{document}